\documentclass[twocolumn,showpacs,preprintnumbers,amsmath,amssymb]{revtex4}

\usepackage[dvips]{graphicx}
\usepackage{dcolumn}
\usepackage{bm}

\begin{document}

\preprint{APS/123-QED}

\title{Multigap superconductivity in sesquicarbides La$_2$C$_3$ and Y$_2$C$_3$}
\author{S. Kuroiwa$^{1}$, Y. Saura$^{1}$, J. Akimitsu$^{1}$, M. Hiraishi$^{2}$, M. Miyazaki$^{2}$, K. H. Satoh$^{2}$, S. Takeshita$^{3}$, and R. Kadono$^{2,3}$ }
\affiliation{
$^1$Department of Physics and Mathematics, Aoyama Gakuin University, Sagamihara, Kanagawa 229-8558, Japan\\
$^2$Department of Materials Structure Science, The Graduate University for Advanced Studies, Tsukuba, Ibaraki 305-0801, Japan\\
$^3$Institute of Materials Structure Science, High Energy Accelerator Research Organization, Tsukuba, Ibaraki 305-0801, Japan
}%

\begin{abstract}

A complex structure of the superconducting order parameter in $Ln_2$C$_3$ ($Ln$ = La, Y) is
demonstrated by muon spin relaxation ($\mu$SR) measurements in their mixed state.  
The muon depolarization rate [$\sigma_{\rm v}(T)$] exhibits a characteristic temperature dependence 
that can be perfectly described by a phenomenological double-gap model for nodeless superconductivity.  While the magnitude of two gaps is similar between La$_2$C$_3$ and Y$_2$C$_3$, a significant difference in the interband coupling between those two cases is clearly 
observed in the behavior of $\sigma_{\rm v}(T)$.

\end{abstract}

\maketitle
The revelation of high-temperature superconductivity in magnesium diboride (MgB$_2$, with critical temperature $T_c\simeq39$ K) 
has stimulated renewed interest in other boride and carbide superconductors as an alternative
path to novel superconductors with an even higher $T_c$ \cite{Nagamatsu:01}.  Sesquicarbides ($Ln_2$C$_3$, $Ln=$ La, Y) are among such compounds reported in early literatures; they exhibit superconductivity at relatively high critical temperatures ($T_{\rm c}\simeq$ 6--11 K) and their $T_{\rm c}$'s strongly depend on carbon composition \cite{Giorgi02,Krupka01,Simon}. 
Recently, we have found a new superconducting phase  in Y$_2$C$_3$ that exhibits a much higher $T_{\rm c}$ ($\sim$ 18 K) comparable with $A$-15 compounds \cite{Amano}.
This discovery has attracted further attention to the relationship between structural details and superconductivity in  sesquicarbide systems.
However, despite various attempts \cite{Francavilla,Cort,Nakane,Kim02}, little is known so far about the details of superconducting order parameters in La$_2$C$_3$ and Y$_2$C$_3$ from a microscopic viewpoint.

A recent study on the temperature dependence of the nuclear spin-lattice relaxation rate in Y$_2$C$_3$ has suggested the occurrence of multiple superconducting gap with $s$-wave symmetry in a sample having $T_{\rm c}$ = 15.7 K \cite{Harada}.  While similar electronic structure would be expected for La$_2$C$_3$ \cite{Kim02}, a report on the specific heat measurement suggests single-gap superconductivity in a specimen with $T_{\rm c}$ $\approx$ 13.4 K \cite{Kim}.
In any case, the real nature of superconductivity in Y$_2$C$_3$ and La$_2$C$_3$, including potential difference between the two systems, still remains largely unclear.

The muon spin rotation ($\mu$SR) technique is a useful microscopic tool for  probing quasiparticle (QP) density of states available for thermal$/$field-induced 
excitation in the mixed state of type II superconductors \cite{Sonier,Kadono}. The muon depolarization rate in the mixed state is predominantly determined by the magnetic penetration depth  ($\lambda$) that is controlled by superfluid density. Since the latter is reduced by the QP excitation, the effective value of  $\lambda$ serves as a monitor of the QP excitation.
In this letter, we present the result of $\mu$SR measurements on polycrystalline samples of La$_2$C$_3$ ($T_{\rm c}$ $\sim$ 11 K) and Y$_2$C$_3$ ($T_{\rm c}$ $\sim$ 15 K), where a clear sign 
of double-gap superconductivity is observed in the temperature dependence of the muon 
depolarization rate.  They also provide the first clear case for
the double-gap model, where the magnitude of coupling between electronic bands responsible for superconductivity is explicitly examined.
Our result establishes a coherent description of multiple band$/$gap superconductivity in this sesquicarbide system.


\par
For the $Ln_2$C$_3$ samples, starting materials were prepared by the arc melting method using a mixture of La$/$Y (99.9 $\%$) and C (graphite, 99.99 $\%$) with stoichiometric composition of sesquicarbide.
The obtained Y-C alloys were placed into a BN cell in a dry box under an argon gas atmosphere, and polycrystalline Y$_2$C$_3$ was synthesized by elevating temperature to 1300 $\sim$ 1400 $^{\rm \circ}$C for 30 min under a high pressure of 5 GPa using a cubic-anvil-type equipment.
For the polycrystalline La$_2$C$_{3}$, the La-C alloys obtained by the arc melting were pressed into pellets in a sealed Ta tube, and sintered at 1000 $^{\rm \circ}$C for 200 h under a high vacuum condition of 3.0 $\times$ 10$^{-5}$ Torr, followed by a slow cooling process to ambient temperature at a rate of 5 $^{\rm \circ}$C$/$h.
\par
The powder x-ray diffraction patterns for both specimens could be indexed as a sesquicarbide phase with the space group of $I{\bar{4}}3d$.
In La$_2$C$_3$, nearly 10\% of LaC$_2$ was observed as a minor phase besides that of 
the sesquicarbide, while Y$_2$C$_3$ was found to be in a single phase. LaC$_2$ behaves as a normal metal above 2 K and only causes a background in the $\mu$SR signal in the superconducting phase. The lattice constants of La$_2$C$_{3}$ and Y$_2$C$_{3}$ were determined to be approximately $a$ = 8.808(5) {\AA} and 8.238(5) {\AA}, respectively, which are in good agreement with those reported previously \cite{Kim,Novokshonov,Mochiku,Wang}. 
Unfortunately, the precise stoichiometry of carbon has not been determined.
Therefore, the chemical composition in this paper refers only to a nominal value.
Heat capacity and ac and dc magnetic susceptibilities were measured using MPMSR2 and PPMS (Quantum Design Co., Ltd.). 
 
\begin{figure}[t]
\begin{center}
\includegraphics[width=1\linewidth]{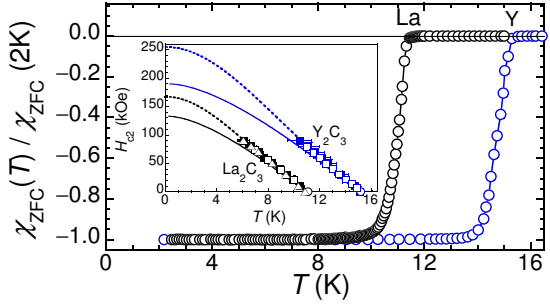}
\caption{\label{Fig1} (Color online) Temperature dependence of dc magnetic susceptibility  at 10 Oe in La$_2$C$_3$ and Y$_2$C$_3$ normalized by the value at 2 K.  Inset shows magnetic field ($H$) vs. temperature ($T$) phase diagram. The triangle, circle, and square symbols indicate data determined by heat capacity and dc and ac magnetic susceptibility measurements, respectively. The open and closed symbols show data obtained from $T$ and $H$-scan, respectively. The solid and dashed curves correspond to the WHH relation and the GL model, respectively.}
\end{center}
\end{figure}

\par

Figure {\ref{Fig1} shows the temperature dependence of dc susceptibility under zero-field cooling 
in La$_2$C$_3$ and Y$_2$C$_3$.
 In both cases, clear diamagnetic signals are observed below 
$T_{\rm c}$ ($\simeq$ 11.2 K and 15.2 K, respectively).
As shown in the inset, $H_{\rm c2}(T)$ exhibits almost linear  dependence on temperature, which differs significantly from the Werthamer-Helfand-Hohenberg (WHH) relation \cite{Werthamer}.
The enhancement of $H_{\rm c2}(T)$ from the WHH prediction is due to the strong electron-phonon coupling rather than the anisotropic Fermi surface or localization effect \cite{Kim}.
We can extract $H_{\rm c2}$(0) without much uncertainty using the Ginzburg-Landau (GL) theory, $H_{\rm c2}(T)$ = $H_{\rm c2}(0)(1-(T/T_{\rm c})^2)/(1+(T/T_{\rm c})^2)$, where $H_{\rm c2}(0)$ = $\Phi_0/2\pi{\xi_{\rm GL}}^2$, $\Phi_0$ is the flux quantum, and $\xi_{\rm GL}$ is the GL-coherence length  \cite{Tinkham}.
The best fit using the above equation yields $H_{\rm c2}$(0) = 167(3) and 256(7) kOe for La$_2$C$_3$ and Y$_2$C$_3$, respectively. 

Conventional $\mu$SR experiment was performed on the M15 beamline of TRIUMF, Canada.
The polycrystalline samples were loaded on a sample holder (a scintillator serving
as a muon veto counter, with a sample dimension of 7 $\times$ 7 mm$^2$) and placed into a He gas-flow cryostat, to which a 100\% spin-polarized muon beam with a  momentum of 29 MeV$/c$ was irradiated to collect $1.5\times10^{7}$ decay positron events for each spectrum
(taking about 1.5 h). 
Each measurement was performed under a field-cooling process to minimize the effect of flux pinning, and field fluctuation was kept within 10$^{-4}$ of the applied field.

Since we can reasonably assume that muons stop randomly on the length scale of the flux-line lattice (FLL), the muon spin precession signal, $\hat{P}(t)$, provides the random sampling of the internal field distribution $B({\bf r})$,
\begin{eqnarray}
\hat{P}(t) 
         & = & \int^{\infty}_{-\infty}n(B)\cos(\gamma_{\mu}Bt+\phi)dB, \\
n(B) & = & \langle{\delta(B-B(\bf r))}\rangle_{\bf r}
\end{eqnarray}
where $\gamma_{\mu}$ is the muon gyromagnetic ratio (= 2$\pi\times13.553$ MHz/kOe), $n(B)$ is the spectral density for the internal field defined as a spatial average $(\langle{\rangle}_{\bf r})$ of the delta function, and $\phi$ is the initial phase of rotation.
These equations indicate that the real amplitude of the Fourier transformed muon spin precession signal corresponds to $n(B)$ (except corrections for additional relaxation due to other origins, see below).
In the case of relatively large magnetic penetration depth ($\lambda\ge3000$ \AA), $n(B)$ can be well-approximated by a simple Gaussian field profile, yielding $\hat{P}(t)\simeq\exp(-\sigma^2t^2/2)\cos(\gamma_{\mu}\bar{B}t+\phi)$, where $\sigma$ = $\gamma_{\mu}\sqrt{\langle{(B-B({\bf r}))^2}\rangle}\propto\lambda^{-2}$ and $\bar{B}\simeq H$ is the mean field.  Here, it must be stressed that 
$\lambda$ is an {\sl effective} magnetic penetration depth susceptible to the quasiparticle excitation.

\begin{figure}[t]
\begin{center}
\includegraphics[width=1\linewidth]{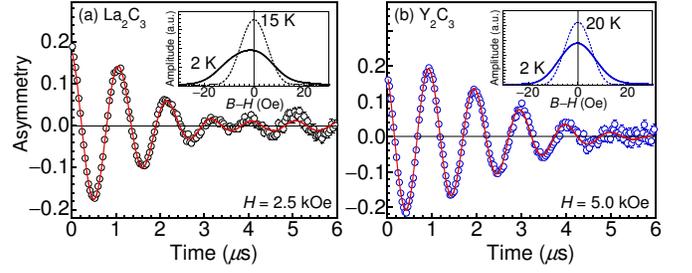}
\caption{\label{Fig2} (Color online) Time revolution of muon-positron decay asymmetry in (a) La$_2$C$_3$ and (b) Y$_2$C$_3$ at 2 K under a transverse field of 2.5 kOe and 5.0 kOe, respectively, displayed in a rotating-reference-frame frequency of (a) 33 MHz and (b) 66.8 MHz. The respective insets show the fast Fourier transform (FFT) at 2 K (solid lines) and above $T_{\rm c}$ (dashed lines).}
\end{center}
\end{figure}

Figure {\ref{Fig2}} shows the time-dependent muon-positron decay asymmetry at 2 K in La$_2$C$_3$ and Y$_2$C$_3$ with their fast Fourier transform (FFT) displayed in the inset.
The FFT spectral linewidth in the normal state ($T>T_{\rm c}$) is determined by the small random local fields from nuclear moments and a limited $\mu$SR time window ($\simeq8$ $\mu$s), while that in the  superconducting state is further broadened by the formation of FLL and associated inhomogeneous local field
distribution [$B({\bf r})$].
The solid curves in the main panels are the best fits of the data in the time domain, assuming two components of the Gaussian damping,
\begin{equation}
A\hat{P}(t)~=~\sum_{i=1}^{2}A_i\exp\left(-\frac{\sigma_i^2t^2}{2}\right){\cos}(\gamma_{\mu}B_it+\phi_i)
\end{equation}
where the $i$-th component refers to the contribution from superconducting ($i$ = 1) and normal ($i$ = 2) phases, $A_i$ is the partial asymmetry ($\sum_iA_i=A$), $\sigma_i$ is the relaxation rate, and $\gamma_\mu B_i$ is the central frequency for the respective components.
The model yields good fits to data, as indicated by the reasonably small values of reduced chi-square: $\chi^2/N_f$ is mostly less than 1.7 for La$_2$C$_3$ and 1.3 for Y$_2$C$_3$, with $N_f$ being the number of degrees of freedom. Considering that $\sigma_2$ represents the relaxation due to the nuclear magnetic moment (i.e., $\sigma_2$ = $\sigma_{\rm n}$), the net relaxation rate in the superconducting state is expressed as $\sigma_1^2 = \sigma_{\rm n}^2 + \sigma_{\rm v}^2$, where the second term comes from $n(B)$ in the FLL state and it is proportional to the superfluid density \cite{Brandt}. 
From the fitting analysis, the superconducting volume fractions [= $A_1/(A_1+A_2$)] at 2 K are estimated to be $\approx$ 0.91 and 0.98 in La$_2$C$_3$ and Y$_2$C$_3$, respectively, where the former value is in good agreement with the fractional yield estimated by the X-ray analysis (with the rest corresponding to LaC$_2$ in the normal
state).

\begin{figure}[t]
\begin{center}
\includegraphics[width=1\linewidth]{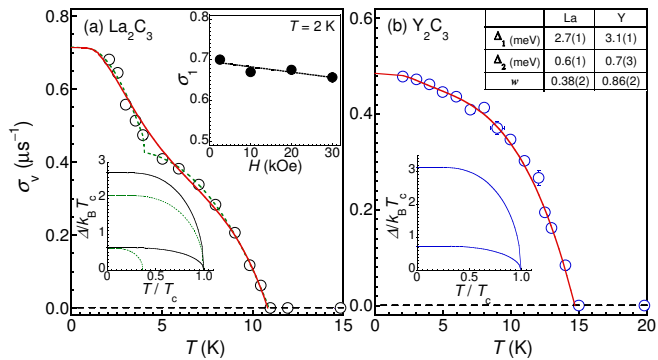}
\caption{\label{Fig3} (Color online) Temperature dependence of muon spin relaxation rate for (a) La$_2$C$_{3}$ at 2.5 kOe and (b) Y$_2$C$_{3}$ at 5.0 kOe. Error bars (not shown) are smaller than the symbol size. Solid and dashed curves indicate the result of fitting analysis using the double-gap model described in the text. Insets show the relaxation rate in the superconducting state ($\sigma_{1}$) as a function of magnetic field for La$_2$C$_{3}$ (a) and the order parameters [$\Delta(T)/k_{\rm B}T_{\rm c}$] for the respective cases. }
\end{center}
\end{figure}

Figure \ref{Fig3} shows the temperature dependence of $\sigma_{\rm v}$ for La$_2$C$_3$ at 2.5 kOe and Y$_2$C$_3$ at 5.0 kOe, where the fields are chosen to be reasonably away from the lower critical field (the field dependence at the relevant field range is shown for La$_2$C$_3$). It is interesting to note that the data of La$_2$C$_3$ exhibit a shoulder-like structure near 4 K, indicating clear deviation from the behavior predicted for single-gap BCS superconductors.
The effect of flux pinning as a possible origin of such structure is ruled out by the fact that $\sigma_{\rm 1}$ at 2 K is almost independent of the applied field below 30 kOe [see the inset of Fig.~\ref{Fig3}(a)].
According to the theories that consider multiband superconductivity \cite{Suhl,Xiang,Golubov}, such an inflection is expected to occur in the systems consisting of two superconducting bands that are weakly coupled.  The behavior of $\sigma_{\rm v}$  below $\sim$4 K is then attributed to the band with a smaller gap energy. 
To our knowledge, this is the first unambiguous example of double-gap superconductivity
with extremely week interband coupling.
\par
Compared with La$_2$C$_3$, no strong anomaly is observed in the case of Y$_2$C$_3$.
However, while the superfluid density (and hence $\sigma_{\rm v}$) is predicted to be virtually independent of temperature for $T/T_{\rm c}$ $\leq$ 0.4 in the single-gap BCS model, a clear variation of $\sigma_{\rm v}$ with temperature is observed below 6 K.
This is another sign that the order parameter of Y$_2$C$_3$ has an anisotropic structure.  
Considering the present result in La$_2$C$_3$ and previous NMR studies \cite{Harada}, 
we can attribute the $T$-dependence of $\sigma_{\rm v}$ also to the double-gap superconductivity.  This is further supported by a recent first-principles calculation \cite{Nishikayama},
where it is suggested that the Fermi surface of Y$_2$C$_3$ consists  of three hole bands (0.12, 0.15, and 0.88$/$eV unit cell spin) and one electron band (2.73$/$eV unit cell spin) that arise mainly from the hybridized orbitals of Y $d$- and C-C antibonding $\pi^{\ast}$-states.
Such large differences in the density of states and Fermi velocities between hole and electron bands might lead to the opening of two superconducting gaps in the different parts of the Fermi surface.
\par
The origin of difference in the temperature dependence of $\sigma_{\rm v}$ between La$_2$C$_3$ and Y$_2$C$_3$ is understood by considering the difference in the interband coupling strength between these two compounds. 
For quantitative discussion, the data in Fig.~\ref{Fig3} are analyzed using a phenomenological double-gap model with $s$-wave symmetry \cite{Bouquet,Ohishi},
\begin{eqnarray}
\sigma(T)&=&\sigma(0)-w{\cdot}\delta\sigma(\Delta_1,T)-(1-w){\cdot}\delta\sigma(\Delta_2,T),  \nonumber \\
\delta\sigma(\Delta,T)&=& \frac{2\sigma(0)}{k_{\rm B}T}\int_0^{\infty}f(\epsilon,T){\cdot}[1-f(\epsilon,T)]d\epsilon, \nonumber \\
f(\epsilon,T) &=& \left(1+e^{\sqrt{\epsilon^2+\Delta(T)^2}/k_{\rm B}T}\right)^{-1},\nonumber 
\end{eqnarray}
where $\Delta_i$ ($i$ = 1 and 2) is the energy gap at $T=0$, $w$ is the relative weight for $i=1$, $k_{\rm B}$ is the Boltzmann constant, $f(\epsilon,T)$ is the Fermi distribution function, and $\Delta(T)$ is the standard BCS gap energy.
The solid curves in Fig.~\ref{Fig3} are the best fit result obtained by using the above double-gap model with the parameters listed in Table {\ref{Table1}}.
For La$_2$C$_3$, a simplified model (dashed curve) assuming two independent superconducting bands was also tested against the data, which turned out to exhibit a slightly better agreement than that described by using the above model [yielding  $2\Delta_1/k_{\rm B}T_{\rm c}=4.5(3)$ and $2\Delta_2/k_{\rm B}T_{\rm c}=1.3(3)$]. This might suggest that the above model may not necessarily be a good approximation for 
the case of weak interband coupling.
 
\begin{table}[t]\renewcommand{\arraystretch}{1.1}\caption{Superconducting properties of La$_2$C$_3$ and Y$_2$C$_3$ determined from the present experiment, where those obtained from the double-gap analysis correspond to the solid curves in Fig.~\ref{Fig3}.}\begin{center} \begin{tabular}{lcccccccc}
  \hline\hline
  ~~~~ & ~~~~~~~~~~{\bf La$_2$C$_3$}~~~~~~~~~~ & ~~~~~{\bf Y$_2$C$_3$}~~  \\
\hline 
  ~Transverse field (kOe) & 2.5 &  5.0 \\
  \hline
  ~$T_{\rm c}$ (K) & 10.9(1) &  14.7(2) \\
  ~$\sigma_{\rm v}$(0) ($\mu$s$^{-1}$)& 0.71(3) & 0.48(2) \\   
  ~$\lambda$(0) (\AA)  & 3800(100) & 4600(100)  \\
  ~$w$ & 0.38(2) & 0.86(2) \\
  ~$\Delta_1$(0) (meV)  & 2.7(1) &  3.1(1)   \\
  ~$\Delta_2$(0) (meV)  & 0.6(1) &  0.7(3)  \\
  ~$2\Delta_1/k_{\rm B}T_{\rm c}$  & 5.6(3) & 4.9(3) \\
  ~$2\Delta_2/k_{\rm B}T_{\rm c}$  & 1.3(3) & 1.1(5)\\
  \hline \hline
 \end{tabular}
 \label{Table1}\end{center}\end{table}

The superconducting parameters deduced from the present experiment are summarized in Table {\ref{Table1}}.
Here, we calculated the magnetic penetration depth $\lambda$(0) using the formula
$\sigma_{\rm v}(0)~[\mu{\rm s}^{-1}]=4.83{\times}10^4(1-H/H_{\rm c2})\times[1+3.9(1-H/H_{\rm c2})^2]^{1/2}/\lambda^{2}(0)$ [nm] \cite{Brandt,Aeppli}.
The gap parameter $2\Delta/k_{\rm B}T_{\rm c}$ of Y$_2$C$_3$ is in reasonable agreement with that
deduced by NMR (i.e., $2\Delta/k_{\rm B}T_{\rm c}$ = 5 and 2) \cite{Harada}, again supporting the present double-gap scenario. 
We also find that $2\Delta/k_{\rm B}T_{\rm c}$ for the two respective bands of La$_2$C$_3$ are comparable with those of Y$_2$C$_3$.
Thus, it appears that the superconductivity of La$_2$C$_3$ and Y$_2$C$_3$ share the common features of strong electron-phonon coupling and $s$-wave symmetry, which is in line with the previous heat capacity results \cite{Kim}.
\par
Provided that there is a significant difference in the interband coupling between La$_2$C$_3$ and
Y$_2$C$_3$, the observed difference in the relative weight ($w$) between two gaps might also be connected with the interband coupling.
Furthermore, considering that the double-gap features tend to be suppressed by the localization (scattering) effect, one might suspect that such a difference in $w$ may arise from that in the quality of 
specimen. In this regard, we have to note that the Y$_2$C$_3$ samples were obtained only in a polycrystalline form using high pressure synthesis and that their short annealing time might have resulted in a quality less than that of La$_2$C$_3$.
At this stage, we presume it unlikely that the present result has been strongly affected by the localization effect, considering that the electronic mean free path measured using the microwave cavity perturbation technique is much longer than $\xi_{\rm GL}$ for the sample prepared under the same condition \cite{Akutagawa}.  However, it would be 
certainly helpful to study the influence of sample quality (and chemical stoichiometry as well) in the future to elucidate the details of the localization effect on the double-gap behavior.
 \par 
Finally, let us point out the noncentrosymmetric effect in superconductivity.
 In the case of a sesquicarbides system with the $I\bar{4}3d$ group symmetry, an asymmetric spin-orbit interaction can be approximated by the Dresselhaus-type interaction.
When the order of magnitude of a superconducting gap is of comparable to that of the spin-orbit band splitting, the original isotropic gap structure is modulated by a magnetic field  to have a point-node, because anisotropic Pauli depairing effect can occur in the specific part of the momentum space \cite{Fujimoto}.
This may lead to unusual field-induced quasiparticle excitation, and a detailed $\mu$SR study on the field dependence of magnetic penetration depth is currently in progress to examine the proposed scenario.

In summary, we performed $\mu$SR experiment on $Ln_2$C$_3$ ($Ln$ = La, Y) to clarify the structure of superconducting order parameter through the temperature dependence of quasiparticle excitation reflected in the muon depolarization rate, $\sigma_{\rm v}(T)$ in the mixed state.
We showed that $\sigma_{\rm v}(T)$ exhibits a characteristic of double-gap in the superconducting order parameter, with a marked variation in the temperature dependence between La and Y compounds that is attributed to the difference in the interband coupling.
The gap parameters for two respective bands were deduced using the phenomenological double-gap model and were found to be comparable between La and Y compounds,
which is consistent with the occurrence of a strong-coupling superconductivity with $s$-wave symmetry in both the systems.

We thank the TRIUMF staff for technical support during the $\mu$SR experiment.
The experiment was partially supported by the KEK-MSL Inter-University Program for Oversea Muon Facilities and the Grant-in-Aid for Scientific Research on Priority Area from the Ministry of
Education, Culture, Sports, Science and Technology of Japan.  One of the authors (S.K.) acknowledges the support of JSPS Research Fellowships.

\end{document}